\documentclass{camera}
\usepackage{graphicx}  
\usepackage{amsfonts,amsmath}

\begin{document}

%
\title{$^9$Be scattering within a four-body continuum-discretized coupled-channels framework}

%
\author{J. Casal, M. Rodr\'iguez-Gallardo \and J. M. Arias}

%
\organization{Departamento de F\'isica At\'omica, Molecular y Nuclear, Facultad de F\'isica, Universidad de Sevilla, Apto. 1065, E-41080 Sevilla, Spain}

\maketitle

\begin{abstract}
The scattering of $^9$Be on $^{208}$Pb is addressed within a four-body Continuum-Discretized Coupled-Channels (CDCC) framework, considering a three-body $\alpha + \alpha + n$ projectile plus a structureless target. The projectile states are generated using the analytical Transformed Harmonic Oscillator (THO) basis in hyperspherical coordinates. Both the elastic and breakup channels are described on the same footing. We find a good agreement between our calculations and the experimental data at beam energies around and below the Coulomb barrier. 
\end{abstract}

%

\section{Introduction}

CDCC calculations~\cite{Austern87} rely on a good theoretical description of the projectile states and on the knowledge, either experimentally or theoretically, of the interaction potential between the projectile fragments and the target. The continuum discretization for two-body projectiles can be achieved using the standard \textit{binning} procedure~\cite{Austern87}. This method was extended for three-body projectiles such as the halo nuclei $^6$He~\cite{MRoGa09} or $^{11}$Li~\cite{Juanpi13}, comprising a core and two valence neutrons. 

An alternative to the \textit{binning} procedure are the pseudo-state (PS) methods, which consist in representing the continuum spectrum of the projectile by the eigenstates of its Hamiltonian in a discrete basis. This method has been successfully applied to reactions induced by two-body~\cite{AMoro09} and three-body projectiles~\cite{MRoGa08}. PS methods can also be applied to reactions induced by three-body projectiles with more than one charged particle~\cite{Descouvemont15,JCasal15}, for which the calculation of actual continuum states is a non-trivial problem~\cite{Nguyen1}.

In this work we describe the scattering of $^9$Be by a $^{208}$Pb target within a four-body CDCC framework. We discretize the $^9$Be spectrum using PS, diagonalizing the three-body Hamiltonian in an analytical THO basis in hyperspherical coordinates (See Refs.~\cite{JCasal13,JCasal14} for details). We calculate the elastic scattering and breakup angular distributions, as well as the total breakup cross section, at different incident energies.

\section{$\boldsymbol{^9}$Be projectile}
Although $^9$Be is a stable nucleus, it has a small binding energy with respect to the two-body and three-body thresholds~\cite{Tilley04}. It shows a Borromean structure $\alpha + \alpha + n$, since none of its binary subsystems $^8$Be or $^5$He is bound. Two-body models for $^9$Be need to assume a $^8{\rm Be} + n$ or $^5{\rm He} + \alpha$ cluster structure, while both configurations are naturally included in a three-body model. For that reason, we study reactions induced by $^9$Be within a four-body CDCC framework, considering a three-body projectile plus a structureless target. 


\section{Results}
Calculations are performed including $^9$Be states $j^\pi=3/2^\pm,5/2^\pm,1/2^\pm$, solving the coupled equations with multipole couplings to all orders. The cutoff energy and number of projectile states are fixed to provide converged results. Details are in Ref.~\cite{JCasal15}.
\subsection{Elastic scattering}
\begin{figure}[t]
\centering
\includegraphics[scale=.3]{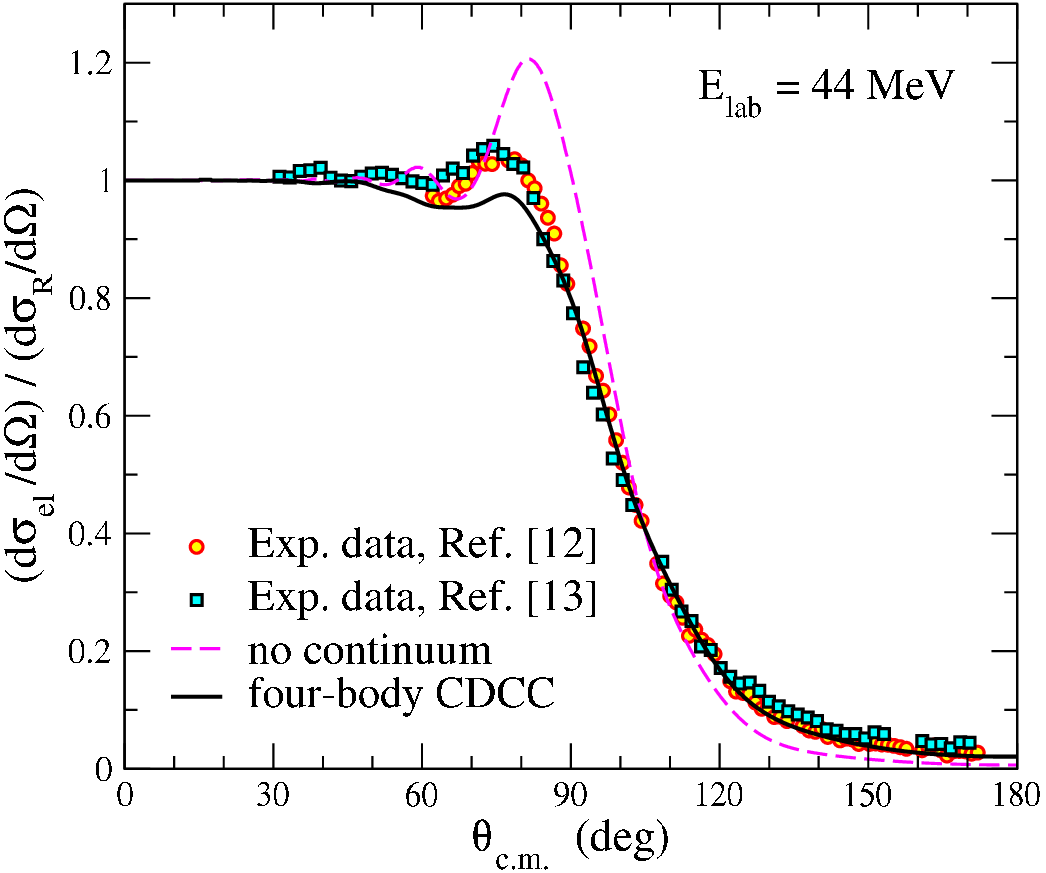} \includegraphics[scale=.3]{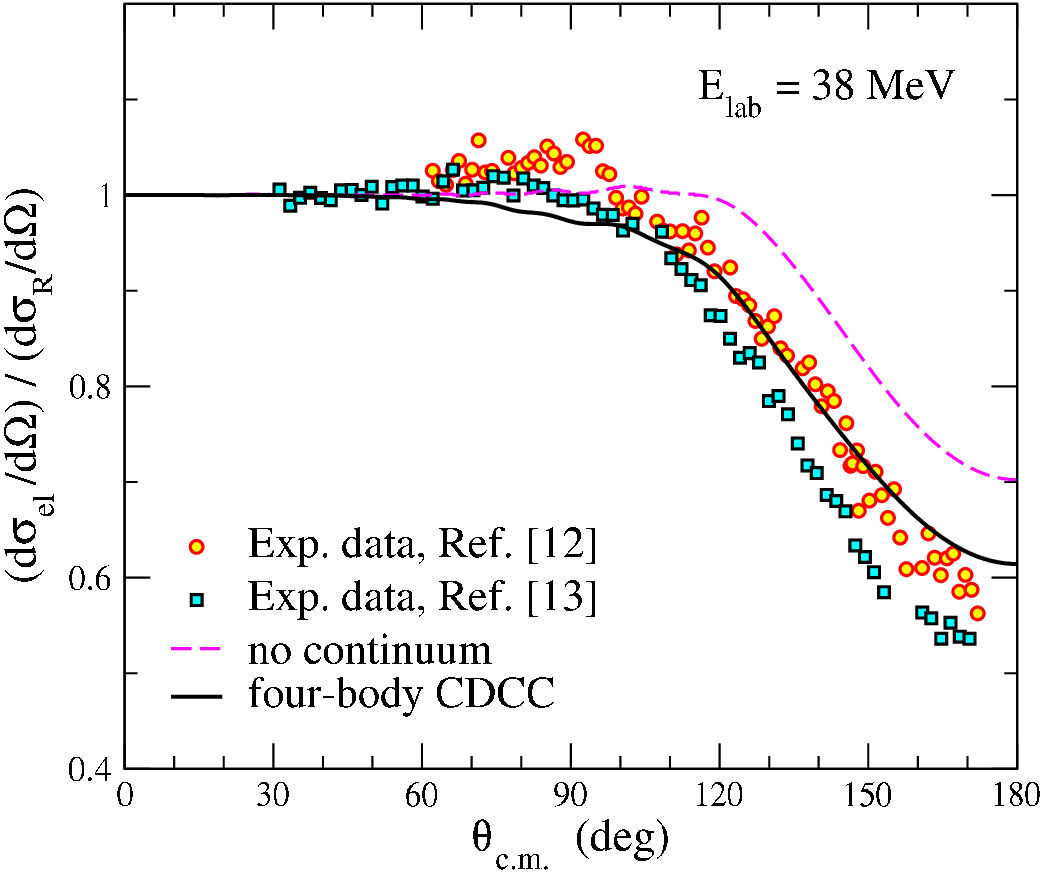}
\caption{$^9$Be + $^{208}$Pb elastic cross section at 44 MeV (left panel) and 38 MeV (right panel).}
\label{fig:elas}
\end{figure}

In Fig.~\ref{fig:elas} we show the elastic cross sections relative to Rutherford at 44 and 38 MeV, i.e.\ around and below the Coulomb barrier, respectively. Dashed lines are calculations including only the $^9$Be ground state and solid lines represent full CDCC calculations. The data are from Refs.~\cite{Woolliscroft04} (circles) and~\cite{Yu10} (squares). We show that the agreement with the data is improved when including the coupling to the continuum, as previously reported in the literature~\cite{Descouvemont15}. Continuum effects are important even at energies well below the Coulomb barrier. Calculations underestimate the experimental data in the nuclear-Coulomb interference region, especially at 44 MeV. This difference may arise from experimental  or theoretical problems, and needs a more exhaustive analysis.

\subsection{Breakup}  
\begin{figure}[t]
\centering
\includegraphics[scale=.3]{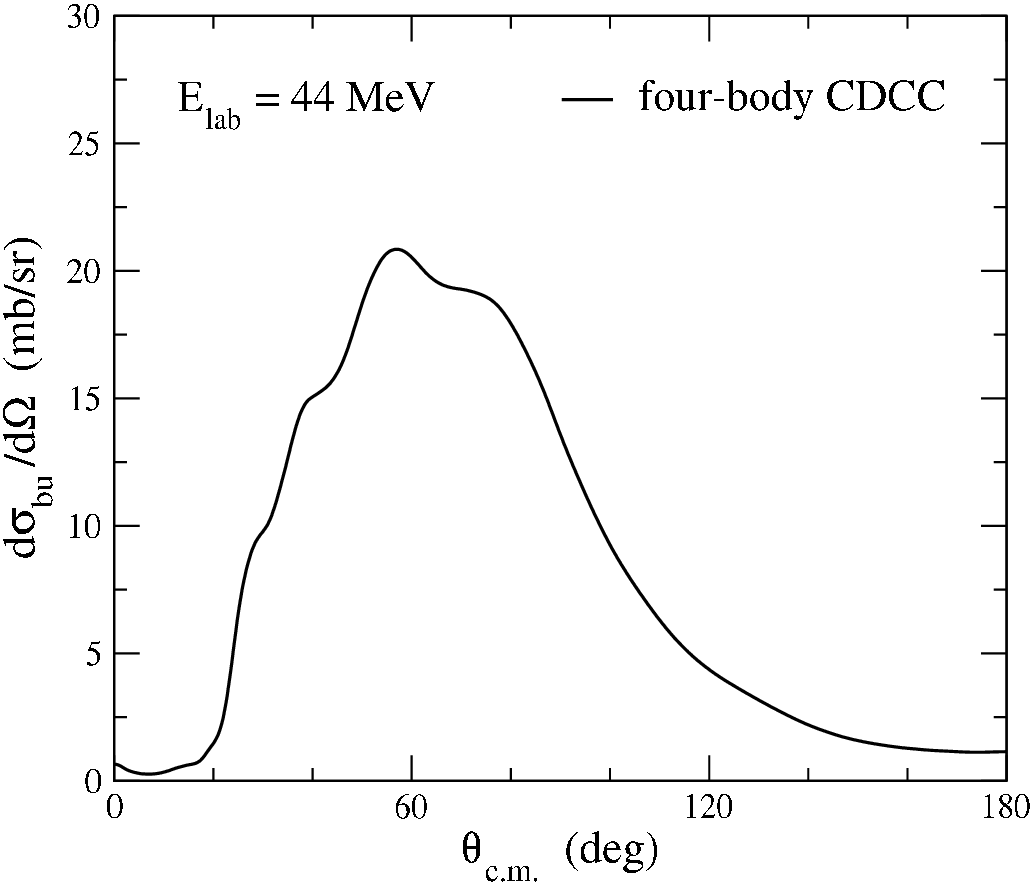}\hspace{5pt} \includegraphics[scale=.3]{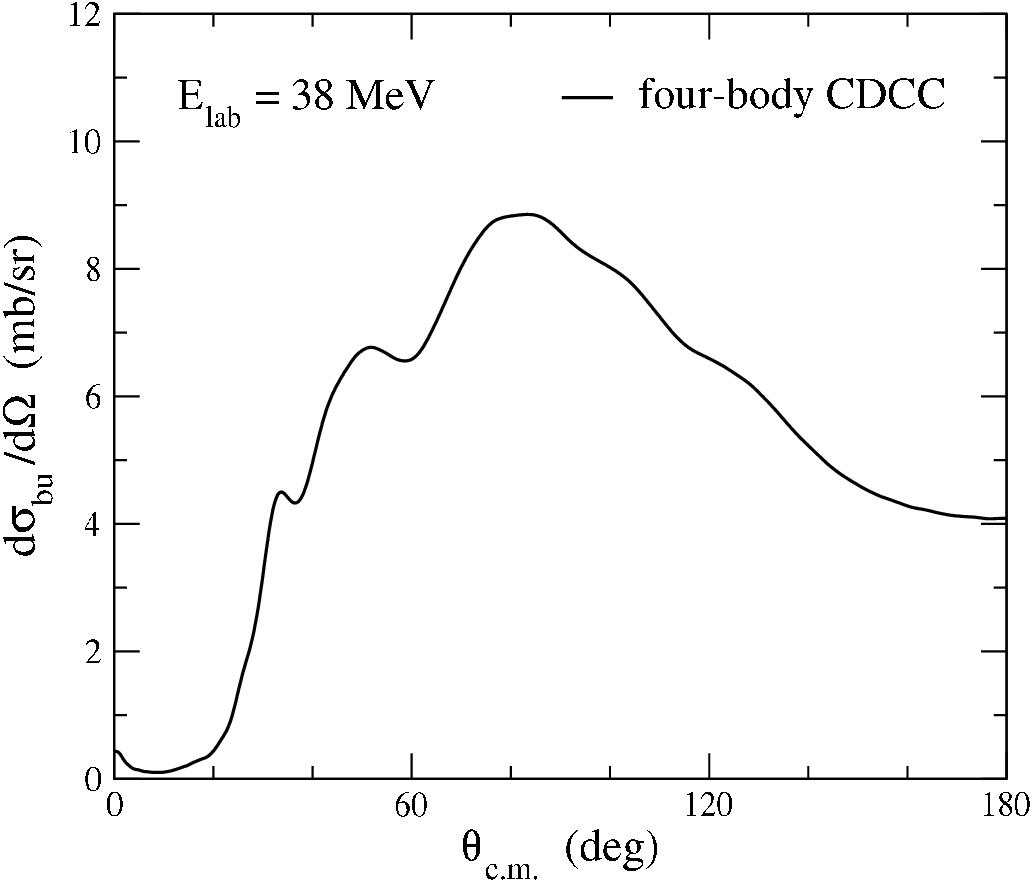}
\caption{$^9$Be + $^{208}$Pb breakup angular distribution at 44 MeV (left panel) and 38 MeV (right panel).}
\label{fig:bu}
\end{figure}

In Fig.~\ref{fig:bu} we show the breakup angular distributions in the center of mass frame at 44 and 38 MeV. The large breakup probability in the nuclear-Coulomb interference region is associated with the reduction of the elastic cross section at the corresponding angles. No data are available in the literature on this quantity. However it could be measured by performing an exclusive Coulomb dissociation experiment. This would allow to asses the validity of the model to describe the elastic and breakup processes on the same footing.
The integrated breakup cross sections at 44 and 38 MeV are 112.0 and 80.5 mb, respectively. This values are in fair agreement with the experimentally reported values of 92.1 and 84.1 mb~\cite{Woolliscroft03}. As expected, the total breakup increases with the incident energy. Breakup effects are important even below the Coulomb barrier, as already shown for the elastic scattering (see Fig.~\ref{fig:elas}).



\section*{Acknowledgements}
This work has been supported by the Spanish Ministerio de Econom\'{\i}a y Competitividad under FIS2014-53448-c2-1-P and FIS2014-51941-P, and by Junta de Andaluc\'{\i}a under group number FQM-160 and Project P11-FQM-7632.
J. Casal acknowledges a FPU grant from the Ministerio de Educaci\'on, Cultura y Deporte, AP2010-3124. M. Rodr\'iguez-Gallardo acknowledges a postdoctoral contract by the VPPI of the Universidad de Sevilla.


%
\end{document}